\documentclass[prl,aps,twocolumn,floatfix,superscriptaddress]{revtex4-1}
\usepackage[latin9]{inputenc}
\usepackage{float}
\usepackage{textcomp}
\usepackage{amsmath}
\usepackage{amssymb}
\usepackage{graphicx}
\usepackage[colorlinks,citecolor=black,urlcolor=blue,bookmarks=false,hypertexnames=true]{hyperref}
\makeatletter
\usepackage{bm}
\usepackage[normalem]{ulem}
\usepackage{tabularx}
\usepackage{wasysym}
\usepackage{color}
\usepackage{array}
\usepackage{multirow}
\usepackage{makecell}
\usepackage{float}
\usepackage{lipsum}
\usepackage{amsfonts}
\usepackage{fancyhdr}
\makeatother

\begin{document}
\title{Multi-orbital model reveals second-order topological insulator in 1H-transition metal dichalcogenide}
\author{Jiang Zeng}
\thanks{Corresponding author: zengjiang@pku.edu.cn}
\affiliation{International Center for Quantum Materials, School of Physics, Peking
University, Beijing, China}
\affiliation{School of Physics and Electronics, Hunan University, Changsha 410082, China}
\author{Haiwen Liu}
\affiliation{Center for Advanced Quantum Studies, Department of Physics, Beijing
Normal University, Beijing 100875, China}
\author{Hua Jiang}
\affiliation{School of Physical Science and Technology, Soochow University, Suzhou
215006, China}
\author{Qing-Feng Sun}
\affiliation{International Center for Quantum Materials, School of Physics, Peking
University, Beijing, China}
\affiliation{Beijing Academy of Quantum Information Sciences, Beijing, China}
\affiliation{CAS Center for Excellence in Topological Quantum Computation, University
of Chinese Academy of Sciences, Beijing, China}

\author{X. C. Xie}
\thanks{Corresponding author: xcxie@pku.edu.cn}
\affiliation{International Center for Quantum Materials, School of Physics, Peking
University, Beijing, China}
\affiliation{Beijing Academy of Quantum Information Sciences, Beijing, China}
\affiliation{CAS Center for Excellence in Topological Quantum Computation, University
of Chinese Academy of Sciences, Beijing, China}

\begin{abstract}
Recently, a new class of second-order topological insulators (SOTIs) characterized by an electronic dipole has been theoretically introduced and proposed to host topological corner states. As a novel topological state, it has been attracting great interest and experimentally realized in artificial systems of various fields of physics based on multi-sublattice models, e.g., breathing kagome lattice. In order to realize such kind of SOTI in natural materials, we proposed a symmetry-faithful multi-orbital model. Then, we reveal several familiar transition metal dichalcogenide (TMD) monolayers as a material family of two-dimensional SOTI with large bulk gaps. The topologically protected corner state with fractional charge is pinned at Fermi level due to the charge neutrality and filling anomaly. Additionally, we propose that the zero-energy corner state preserves in the heterostructure composed of a topological nontrivial flake embedded in a trivial material. The novel second-order corner states in familiar TMD materials hold promise for revealing unexpected quantum properties and applications.
\end{abstract}

\maketitle
Topological insulators are materials with gapped band structure characterized
by quantized topological invariants that are defined with respect
to the symmetries of their bulk Hamiltonian~\cite{2qi2011topological,1hasan2010colloquium}.
In a $d$-dimensional ($d$D) topological insulator, a topologically
non-trivial bulk band structure implies the existence of ($d-1$)D
boundary states. Instead, a $d$D second-order topological insulator
(SOTI) exhibits ($d-2$)D topological states~\cite{4langbehn2017reflection,3benalcazar2017quantized,5ezawa2018higher,15s_schindler2018higher,20s_benalcazar2019quantization,23wang2019higher,24sheng2019two,25ren2020engineering,tang2019comprehensive,hu2018moire,jolad2009testing,lee2020two,park2019higher,liu2021higher}.
For example, there are symmetry protected corner states with localized fractional charge in a 2D SOTI~\cite{3benalcazar2017quantized,5ezawa2018higher,20s_benalcazar2019quantization,23wang2019higher,24sheng2019two,25ren2020engineering,tang2019comprehensive,park2019higher,liu2021higher}. In 2017, the concept of higher-order topological insulators is introduced and characterized by quantized multipole~\cite{3benalcazar2017quantized}. SOTIs and their corner states are investigated in systems with electronic quadruples~\cite{23wang2019higher,24sheng2019two,25ren2020engineering,lee2020two,park2019higher,liu2021higher}. Further in 2018, Ezawa further proposed that the electronic dipole could also induce second-order corner states in a breathing kagome model~\cite{5ezawa2018higher,20s_benalcazar2019quantization, ezawa2018minimal}. This new kind of second-order corner states have been experimentally realized via artificially designing metamaterials in various fields of physics~\cite{6ni2019observation,7imhof2018topolectrical,8noh2018topological,9serra2018observation,11peterson2018quantized,12peterson2020fractional,13kempkes2019robust,14xue2019acoustic,16zhang2019second,17el2019corner,18mittal2019photonic,21xie2019visualization,22fan2019elastic,26cerjan2020observation,yang2020gapped}. 

It is charming to search SOTIs in natural and stable materials, especially the ones have been fabricated through mature technology of high-quality, for further study and application~\cite{33m_li2021printable}. However, experimental demonstration of the existing single-orbital and multi-sublattice models, e.g., the breathing kagome model, in natural electronic materials is still lacking~\cite{13kempkes2019robust}. It is well known that the electronic bands of a material are usually contributed from multi-orbital, due to the degeneracy nature of the atomic orbitals and the hybridization between them. A multi-orbital model would provide a better guidance for realizing intriguing physics in natural materials~\cite{42zeng2021realistic}.

Here, we construct a novel multi-orbital
model to reveal several familiar transition metal dichalcogenides (TMDs)
as a  material family of 2D SOTI with a nontrivial
bulk electronic dipole. The multi-orbital model proposed here shares similarity to the multi-sublattice breathing kagome model~\cite{5ezawa2018higher,20s_benalcazar2019quantization}  and has special advantages on materials realization. Our first-principles calculations and theoretical analysis show that the 1H-MX$_{2}$ monolayers with
M = (W, Mo) and X = (Te, Se, S) are 2D SOTIs with nontrivial   electronic dipoles $\boldsymbol{P}=(\frac{1}{3},\frac{2}{3})$ and large
band gaps about 2 eV, while some other insulators, for example, 1H-TiS$_{2}$,
are topologically trivial. In the topologically nontrivial phase,
the Wannier center of the occupied state locates on neither the M
nor the X atom sites. The mismatch between the Wannier center of
electron and the atom sites refers to an electronic dipole that is
quantized and protected by the $C_{3}$ rotation symmetry. Using a topological nontrivial MoS$_2$ flake as a typical example, our calculations and analysis demonstrate the in-gap corner state with localized fractional $-\frac{1}{3}|e|$ charges. Additionally, the zero-energy corner state can be protected from edge deformation and environmental implication
in the heterostructure composed of a MoS$_{2}$ flake embedded in
a trivial TiS$_{2}$ monolayer. 

Monolayers TMD-MX$_{2}$ with M = (W, Mo) and X = (Te, Se, S)-possess
a variety of polytypic structures. The most-studied 1H structure has
the $D_{3h}$ point-group symmetry and is a sandwich of three planes
of 2D hexagonally packed atoms, X-M-X, as shown in Fig. 1(a). It has
been known that the 1H structure in MX$_{2}$ is typically stable
in free-standing conditions, which is the subject of our work. These
1H-MX$_{2}$ materials have been experimentally fabricated of high
quality~\cite{30m_chhowalla2013chemistry,33m_li2021printable}.

\begin{figure}
\includegraphics[width=1\columnwidth]{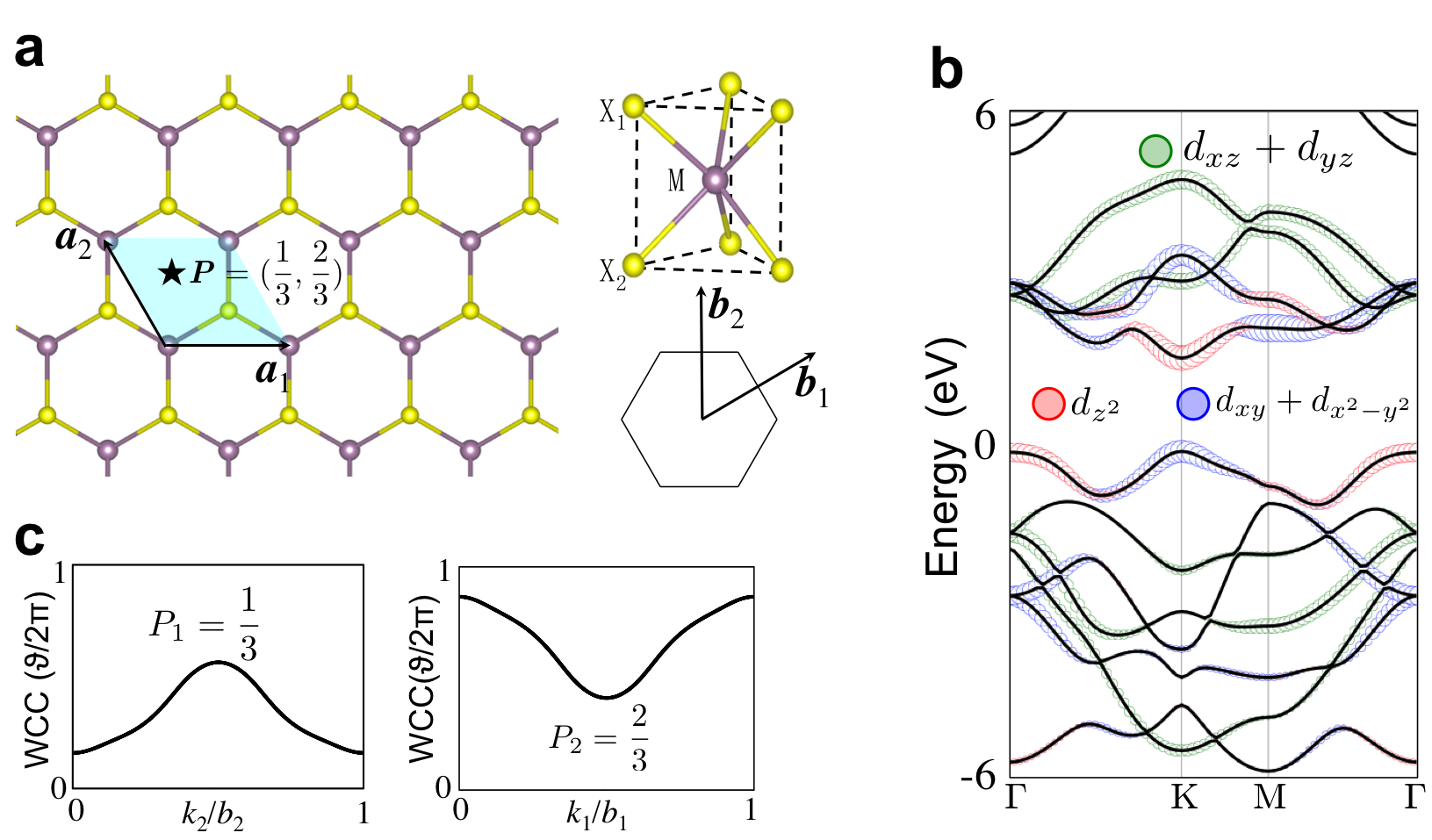} \caption{ Atomic structure and electronic structure of a
1H-MX$_{2}$ monolayer. M stands for (W, Mo) and X stands for (Te, Se, S). \textbf{a}.
Atomic structure of a 1H-MX$_{2}$ monolayer. The arrows $\boldsymbol{a}_{1}$
and $\boldsymbol{a}_{2}$ ($\boldsymbol{b}_{1}$ and $\boldsymbol{b}_{2}$) are the two (reciprocal) lattice vectors. One rhombic unit cell is colored in cyan. The star marks the Wannier charge located at the
hollow site in the unit cell, corresponding to an electronic dipole
$\boldsymbol{P}=(\frac{1}{3},\frac{2}{3})$. The black hexagon in
the lower right corner is the first Brillouin zone. \textbf{b}. Orbital
projected band structure of a 1H-MoS$_{2}$ monolayer. Colored circles
represent contributions from different M-$d$-orbitals. The Fermi
energy $E_{\text{{F}}}$ is set to be zero as a reference. \textbf{c}.
Wilson loop of WCC of the highest valence band showing the calculated
 electronic dipole $\boldsymbol{P}=(\frac{1}{3},\frac{2}{3})$.}
\label{fig1} 
\end{figure}

The electronic structures of various 1H-MX$_{2}$ monolayers were
obtained by first-principles calculations. Figure 1(b) shows a typical
band structure of 1H-MX$_{2}$ using a 1H-MoS$_{2}$ monolayer as
an example, and the results of the other five compounds are shown in Fig.
S1. The 1H-MoS$_{2}$ is an insulator with a fundamental gap of about
2 eV. From early theoretical studies~\cite{35liu2013three}, we
know that the Bloch states of a MoS$_{2}$ monolayer near the band
edges for both conduction and valence bands mostly consist of Mo-$d$-orbitals
with no hybridization between the $d_{z^{2}}$, $d_{xy}$, $d_{x^{2}-y^{2}}$-orbitals
and $d_{xz}$, $d_{yz}$-orbitals, which is explicitly shown in Fig.
1(b). With respect to the symmetry consideration, it is reasonable
to construct a multi-orbital tight-binding model of monolayer MX$_{2}$
using the minimal set of M-$d_{z^{2}}$ , $d_{xy}$, and $d_{x^{2}-y^{2}}$
orbitals as basis~\cite{35liu2013three}. Here we construct a simplified while symmetry faithful
Hamiltonian $H_{\text{s}}$ as 
\begin{align}
H_{\text{s}} & =\sum_{\boldsymbol{r};i;j;k}\frac{t_{23}+t_{32}\pm(t_{23}-t_{32})\epsilon_{ijk}}{2}d_{\boldsymbol{r},i}^{\dagger}d_{\boldsymbol{r}\pm\boldsymbol{a}_{k},j}|\epsilon_{ijk}| \nonumber \\
& +t_{\text{E}}\sum_{\boldsymbol{r};i;j}d_{\boldsymbol{r},i}^{\dagger}d_{\boldsymbol{r},j},
\end{align}
where the hopping parameters for different 1H-MX$_{2}$ monolayers
are listed in Table I. It notes that the indirect interaction mediated
by the X atoms is reflected in the difference between $t_{23}$ and
$t_{32}$, which breaks the inversion symmetry. Figure 2(a) is a
schematic diagram of the above multi-orbital model, which shares similarity to the multi-sublattice breathing kagome model~\cite{5ezawa2018higher,20s_benalcazar2019quantization}. See more details
in the Supplemental Material.

\begin{figure}
\includegraphics[width=1\columnwidth]{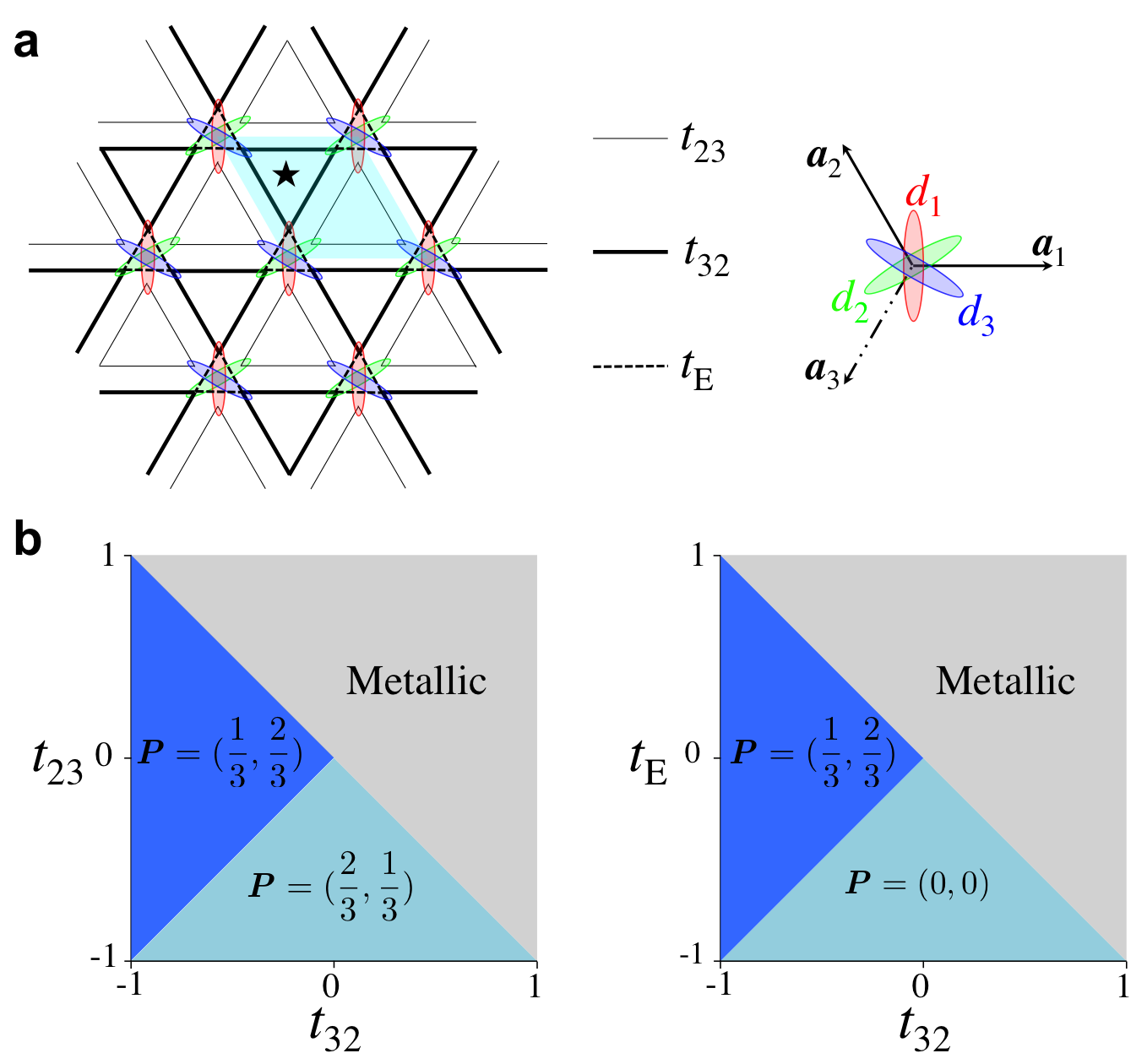} \caption{Schematic lattice diagram and phase diagram of the simplified multi-orbital
model Eq. (1). In (\textbf{a}), the three ellipses in different colors
represent three orbitals on the same site. The thicker and thinner
lines present two different kinds of nearest neighbor hoppings. The dashed lines present the crystal field effects. The
star marks the Wannier charge located at the center of the thicker triangle,
corresponding to  an electronic dipole $\boldsymbol{P}=(\frac{1}{3},\frac{2}{3})$.
Topological phase diagrams of the multi-orbital model for the (\textbf{b})  $t_{\text{E}}=0$ and (\textbf{c}) $t_{23}=0$ cases. }
\label{fig2} 
\end{figure}

The  electronic dipole can be calculated as~\cite{38alexandradinata2014wilson,39vanderbilt1993electric,41fang2012bulk}
\begin{equation}
\boldsymbol{P}=\frac{{1}}{S}\int_{BZ}\text{{Tr}(}\bold{A})d^{2}\bold{k}\text{, }
\end{equation}
where $\bold{A}=-i\langle\Psi|\partial_{\bold{k}}|\Psi\rangle$ is
the Berry connection for all the valence bands, $S$ is the area of
the Brillouin zone, and the integration is over the first BZ. The
two elements $P_{1}$ and $P_{2}$ are actually the average values
of Wannier charge center (WCC) along the two reciprocal lattice vectors
$\boldsymbol{b}{}_{2}$ and $\boldsymbol{b}{}_{1}$, respectively,
with the values module 1 confined in the range of {[}0,1)~\cite{38alexandradinata2014wilson,39vanderbilt1993electric,41fang2012bulk}. The location of Wannier center in real space is $P_{1}\boldsymbol{a}_1+P_{2}\boldsymbol{a}_2$~\cite{38alexandradinata2014wilson,39vanderbilt1993electric,41fang2012bulk}.
For the multi-orbital model here, the   electronic dipole
$\boldsymbol{P}$ is determined by the WCC of the lowest band that
is also the highest valence band and decoupled from other bands. 

Since a 1H-MX$_{2}$ structure shown in Fig. 1(a) has the $D_{3h}$
point-group symmetry, the  electronic dipole must be quantized
as $\boldsymbol{P}=(0,0)$, $(\frac{2}{3},\frac{1}{3})$, or $(\frac{1}{3},\frac{2}{3})$~\cite{41fang2012bulk},
corresponding to a Wannier charge centered on the M site, X site, or hollow
site, respectively. The calculated   electronic dipole is $\boldsymbol{P}=(\frac{1}{3},\frac{2}{3})$
for a 1H-MoS$_{2}$ monolayer corresponding to a Wannier charge located
on the hollow site. The Wilson loop of WCC of the highest valence
band along the two reciprocal lattice vectors is shown in Fig. 1(c).
It is a kind of topologically nontrivial polarization when the Wannier
charge dislocates from the M or X atom sites. It notes that the mismatch
between the Wannier center and the atom site is gauge invariant, though
the calculated value $\boldsymbol{P}$ depends on the choice of the
unit cell~\cite{41fang2012bulk,38alexandradinata2014wilson,39vanderbilt1993electric,5ezawa2018higher}.
Figure 2(a) provides
an intuitive diagram of  the multi-orbital model showing the Wannier charge locates at the center
of the thicker triangle formed by larger hoppings ($|t_{32}|$). In
Fig. 2(b), we plot phase diagrams of the multi-orbital
model of Eq. (1). It shows that the system is nontrivial with $\boldsymbol{P}=(\frac{1}{3},\frac{2}{3})$
when the $t_{32}$ has a negative value and its amplitude is larger
than other hopping parameters. The first-principles calculations and
combined theoretical analysis show that all the six 1H-MX$_{2}$
monolayers with M = (W, Mo) and X = (Te, Se, S) are topologically
nontrivial with $\boldsymbol{P}=(\frac{1}{3},\frac{2}{3})$. This
topologically nontrivial polarization is expected to produce the zero-energy
boundary state at the corner and the filling anomaly due to the coexistence
of the $C_{3}$ rotation symmetry and the charge neutrality~\cite{41fang2012bulk,20s_benalcazar2019quantization,5ezawa2018higher}.

\begin{table}
\caption{Hopping parameters in unit of eV for the six 1H-MX$_{2}$ monolayers
and calculated electronic dipole $\boldsymbol{P}$. }

\centering{}%
\begin{tabular}{ccccccc}
\hline 
 & $\text{MoS}_{2}$  & $\text{WS}_{2}$  & $\text{MoSe}_{2}$  & $\text{WSe}_{2}$  & $\text{MoTe}_{2}$  & $\text{WTe}_{2}$\tabularnewline
\hline 
\hline 
$t_{\text{E}}$  & -0.353 & -0.382 & -0.382 & -0.412 & -0.456 & -0.499\tabularnewline
$t_{32}$  & -0.922  & -1.159  & -0.788  & -0.987  & -0.626  & -0.772\tabularnewline
$t_{23}$  & 0.122  & 0.210 & 0.065  & 0.139 & 0.022  & 0.099\tabularnewline
$\boldsymbol{P}$ & $(\frac{1}{3},\frac{2}{3})$  & $(\frac{1}{3},\frac{2}{3})$  & $(\frac{1}{3},\frac{2}{3})$  & $(\frac{1}{3},\frac{2}{3})$  & $(\frac{1}{3},\frac{2}{3})$  & ($\frac{1}{3},\frac{2}{3}$)\tabularnewline
\hline 
\end{tabular}
\end{table}

\begin{figure}
\includegraphics[width=1\columnwidth]{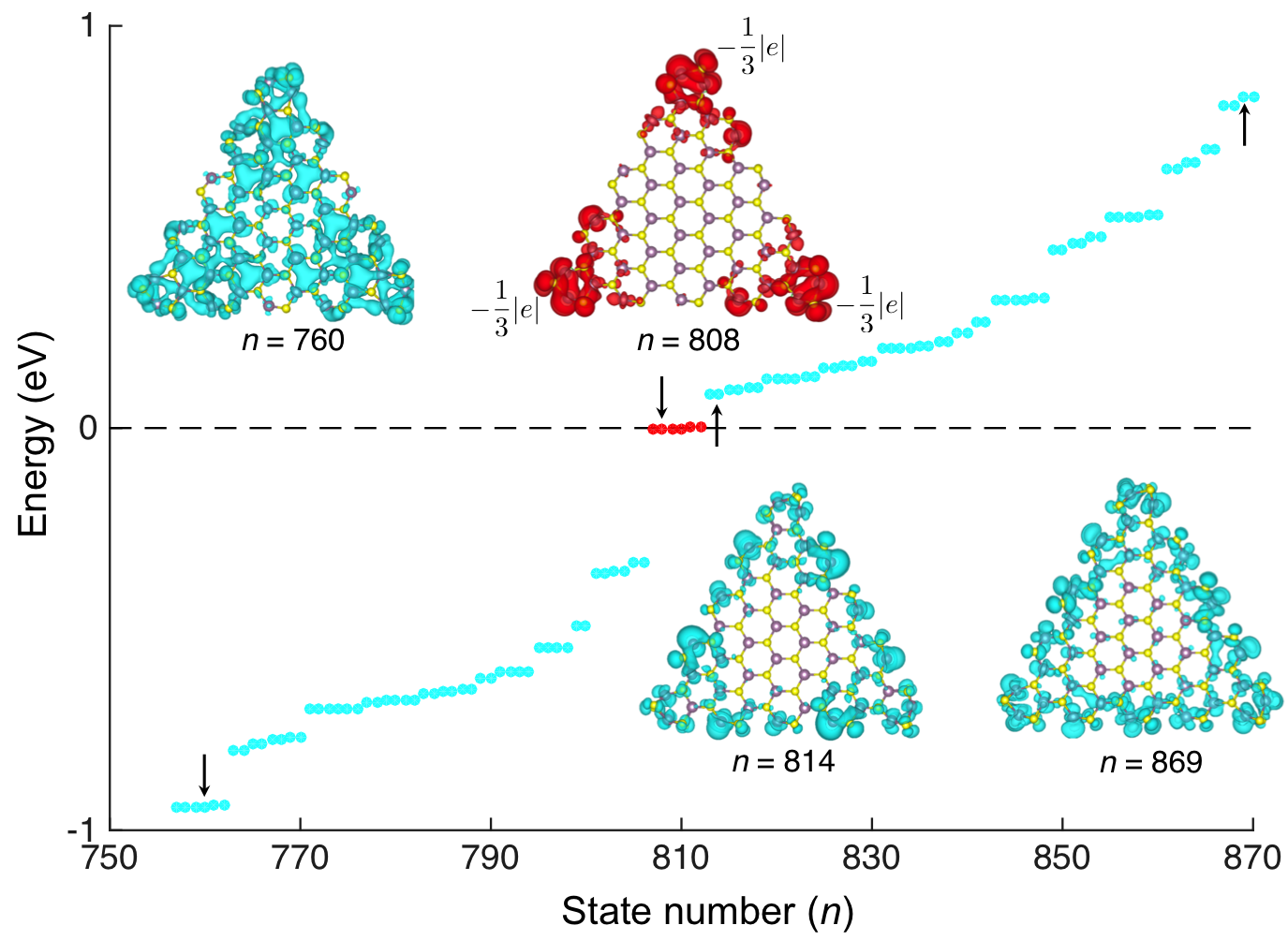} \caption{Energy spectrum of a triangular MoS$_{2}$ flake with armchair edges.
The red dots on the Fermi level represents the 6 in-gap corner states.
Other bulk and edge states are colored in cyan. The atomic structure
and the charge distribution of the corner state $n=808$, bulk state
$n=760$, and edge states $n=814$ and $869$ are shown in the inset.
For the corner state $n=808$, the electron is localized and equally
distributed on the three corners with $-\frac{1}{3}|e|$ charge on
each corner.}
\label{fig3} 
\end{figure}

To explicitly show the zero-energy corner state, a triangular flake
of 1H-MoS$_{2}$ with armchair edges is constructed, as shown in Fig.
3. There are 45 unit cells in the triangular 1H-MoS$_{2}$ flake and
the length of an edge is five hexagons. The triangular shape is
chosen to keep the $C_{3}$ rotation symmetry which is important for
the degeneracy of the corner states. It notes that the appearance of the corner state is sensitive to the
choice of the edge geometry for a topological system protected by
spatial symmetry. Previous works have shown that a zigzag edge of
MoS$_{2}$ has metallic edge states while an armchair edge is insulating~\cite{36cui2017contrasting,37bollinger2001one}. It can be understood in terms of electronic polarization. Since polarization $\boldsymbol{P}$ is perpendicular to the zigzag direction, metallic edge states are expected due to the charge accumulation at the zigzag edge. In contrast, no such charge accumulation and metallic states at the armchair edges that are parallel to $\boldsymbol{P}$. 
A flake with insulating armchair edges is a better choice for the observation
of in-gap corner states.

The first-principles calculated electronic spectrum of the 1H-MoS$_{2}$
flake is shown in Fig. 3. There are 6 corner states at the Fermi level
with 3 corner states for each spin. In our calculations, both spin
degeneracy and spin-orbit coupling are taken into consideration. It
notes that the spin-orbit coupling does not split the spin degeneracy
of the corner states since time-reversal symmetry preserves. As an
example, the charge distribution of the corner state numbered as $n=808$
is shown in red color. The charge distribution of bulk state $n=760$
, edge states $n=814$, and $n=869$ are presented in cyan color for
comparison. For a corner state, one electron is equally distributed
on the three corners with $-\frac{1}{3}|e|$ charges on each corner.
By counting the number of electrons in a charge-neutral flake, the
6 corner states are occupied by 4 electrons at the Fermi level. When the corner states are unoccupied as a result of filling
anomaly~\cite{20s_benalcazar2019quantization}, there will be $\frac{4}{3}|e|$ charges at each corner. See more detailed information in the Supplemental Material.

\begin{figure}
\includegraphics[width=1\columnwidth]{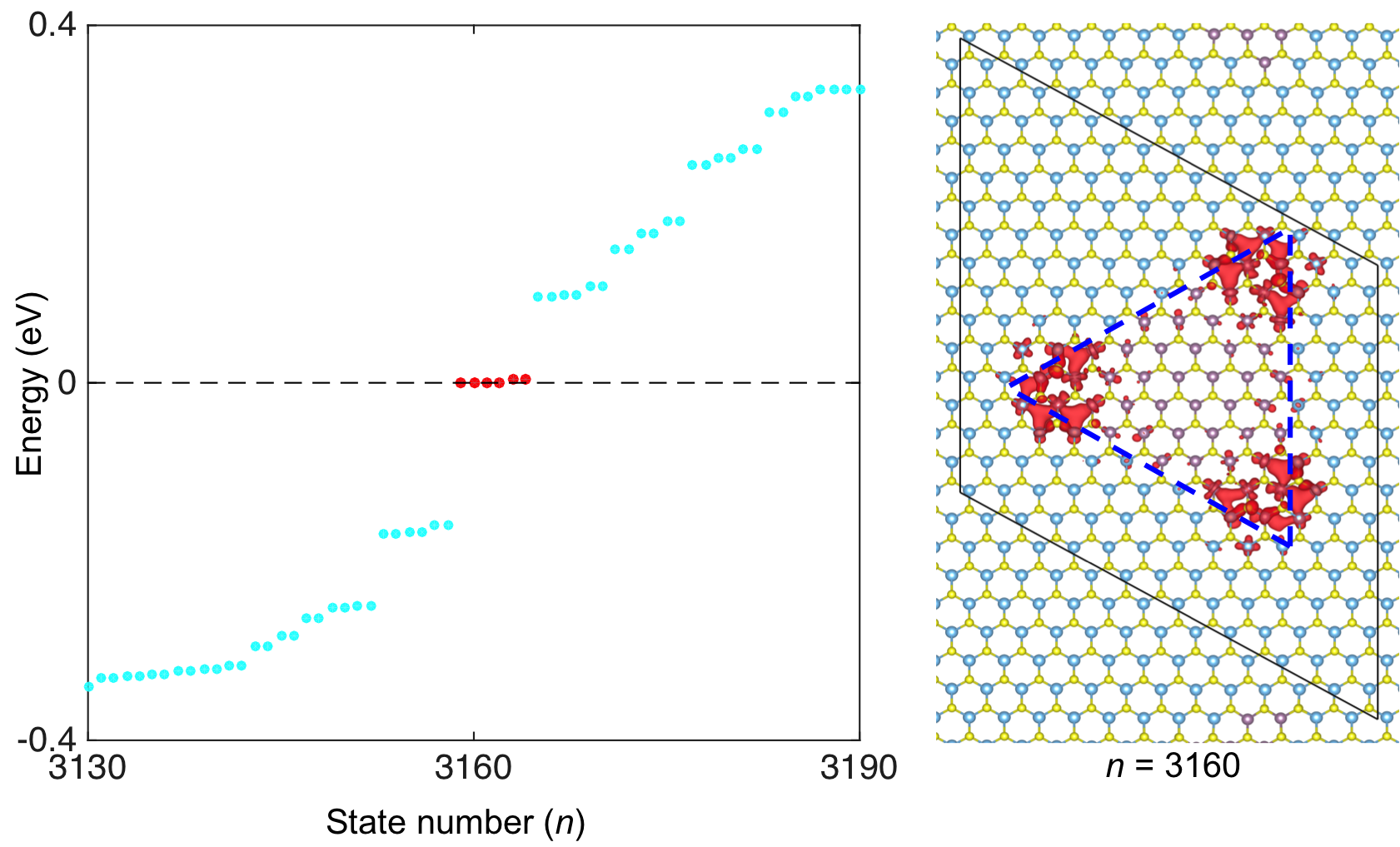} \caption{Energy spectrum of a heterostructure composed of a triangular MoS$_{2}$
flake in a TiS$_{2}$ monolayer. In (\textbf{a}), the red dots on
the Fermi level represent the 6 in-gap corner states. \textbf{b}.
The charge distribution of corner state $n=3160$. The electron is
localized and equally distributed on the three corners with $-\frac{1}{3}|e|$
charges on each corner. }
\label{fig4} 
\end{figure}

Though the corner state is symmetry protected, the degeneracy of the
corner state and the fractional charge nature may deviate from the
ideal case when the $C_{3}$ rotation symmetry is destroyed via edge
deformation or external influence~\cite{6ni2019observation,20s_benalcazar2019quantization,13kempkes2019robust,12peterson2020fractional,11peterson2018quantized}. It is expected that edges and corners
can be protected via embedding the topological 1H-MoS$_{2}$ flake
in a trivial material as a heterostructure. Figure 4 shows a typical
heterostructure of a triangular 1H-MoS$_{2}$ flake in a 1H-TiS$_{2}$
monolayer. It notes that the two materials share similar structures
as well as lattice parameters. In contrast, the 1H-TiS$_{2}$ monolayer
is a topologically trivial insulator because it has two fewer valence
electrons per unit cell less than that of the 1H-MoS$_{2}$ monolayer~\cite{30m_chhowalla2013chemistry,33m_li2021printable}.
We further checked that the MoS$_{2}$/TiS$_{2}$ lateral heterostructure
is insulating at the armchair boundary. Thus, a 1H-TiS$_{2}$ monolayer
provides a perfect platform to protect a topological 1H-MoS$_{2}$
flake and its in-gap corner states. As shown in Fig. 4, the 6 corner
states and their fractional charge nature preserve in the heterostructure.
The 6 corner states are occupied by 4 electrons at the Fermi level, which is the same to the freestanding flake case.

We reveal several familiar TMD monolayers
as a realistic material family of 2D SOTIs. The topologically protected
corner state is pinned at the Fermi level due to the charge neutrality and filling anomaly.
A multi-orbital model is proposed to reveal these TMDs as SOTIs. Additionally,
we propose that the zero-energy corner state of a topologically nontrivial
flake can be further protected from edge deformation and environmental
implication via embedding it in a trivial material using the MoS$_{2}$/TiS$_{2}$
heterostructure as a typical example. The novel second-order corner state in familiar TMD materials hold promise for revealing unexpected quantum properties and applications.

The authors thank Wei Qin and Maoyuan Wang for helpful discussions. This work was supported by the National Basic Research Program of China (2015CB921102 and 2019YFA0308403), the National Natural Science Foundation of China (11674028 and 11822407), the Strategic Priority Research Program of Chinese Academy of Sciences (Grant No. XDB28000000), and China Postdoctoral Science Foundation (2020M670011).

\renewcommand\refname{Reference}
\bibliographystyle{apsrev4-2}
\bibliography{Reference}

\end{document}